Diffusion- Migration-Reaction-Limited Aggregation of Point Defects: A Model for the Stability of Oxide-layers on Metals and Its Breakdown.

K. Ragavendran[a], Bosco Emmanuel[b]

Institute of Mathematical Sciences, Chennai 600113, India

**Abstract**:

A thin oxide layer protects metals from electrochemical corrosion and hence the stability of this oxide layer is crucial for corrosion resistance of metals. The dynamics of cationic and anionic point defects which are injected into this oxide layer at the metal-oxide-layer interface and the oxide-layer-environment interface during electrochemical processes determine the stability of this oxide-layer.

The point defect model originally advanced by Digby Macdonald and perfected into a correct and complete theory by Bosco Emmanuel provides a concrete basis for investigating this stability question. We have formulated a system of 3 coupled differential equations (2 PDE's and one ODE) with appropriate initial and boundary conditions that include the defect injection reactions. This system of equations recognize the diffusion, migration and clustering of the point defects in the oxide-layer. The defect clustering is modelled using the Avrami Theorem and the area-volume laws for random agglomerates.

This research is expected to provide insights into the criticality associated with the oxide-layer breakdown and the periodic, aperiodic and chaotic oscillations in the electrical current / potential which signal the breakdown of the passive oxide-layer. A large body of experimental data is waiting to be understood within a frame-work such as the present. Some initial results from the model are also reported.

[a] Visiting Associate from Kalasalingam Academy of Research and Education, Krishnankoil 626126, India
[b] Corresponding Author:  boscoemmanuel@gmail.com, bosco@imsc.res.in

**Nomenclature and the symbols used in the governing equations:**

| Quantity | Dimension in SI units | Comments |
|---|---|---|
| $C_A^o$ | $\dfrac{mole}{m^3}$ | Initial concentration of the anion defect |
| $C_C^o$ | $\dfrac{mole}{m^3}$ | Initial concentration of the cation defect |
| $R_o$ | $m$ | Initial radius of the void |
| $L$ | $m$ | Thickness of the metal oxide |
| $t$ | $s$ | Time |
| $k_{vf}$ | $\dfrac{m^4}{mole * s}$ | Rate constant of the forward reaction (void growth) |
| $k_{vb}$ | $\dfrac{mole}{m^2 * s}$ | Rate constant of the reverse reaction (void shrinkage) |
| $No$ | $\dfrac{1}{m^3}$ | Number of nucleating sites in the metal oxide film |
| $V_m$ | $\dfrac{m^3}{mole}$ | Molar volume |
| $D_A$ | $\dfrac{m^2}{s}$ | Diffusion coefficient of the anion defect |
| $D_C$ | $\dfrac{m^2}{s}$ | Diffusion coefficient of the cation defect |
| $D_{av}$ | $\dfrac{m^2}{s}$ | Average Diffusion coefficient |
| $\varepsilon$ | $\dfrac{V}{m}$ | Electric Field |
| $\varepsilon_p$ | Dimensionless | Porosity |
| $g^*$ | $V^{-1}$ | $\dfrac{F}{RT}$ |
| $gam$ | Dimensionless | Tortuosity set at 2 |
| $G^{**}$ | Dimensionless | Oxidation state of cation in solution set at 2 |
| $\chi^{***}$ | Dimensionless | Oxidation state of cation in metal oxide set at 2 |
| $t_{RA}$ | Dimensionless | $\dfrac{2k_{vf}\chi\pi L C_C^o Nop}{D}$ |
| $t_{RC}$ | Dimensionless | $\dfrac{4k_{vf}\pi L C_A^o Nop}{D}$ |
| $K_G$ | Dimensionless | $\left(\dfrac{\vartheta_m L}{D}\right) * k_{vf} C_A^o C_C^o$ |
| $Z_A$ | Dimensionless | charge of the anion defect set at $+2$ |
| $Z_C$ | Dimensionless | charge of the cation defect set at $-2$ |
| $K_A$ | Dimensionless | $\dfrac{Z_A \varepsilon L}{\dfrac{RT}{F}}$ |
| $K_C$ | Dimensionless | $\dfrac{Z_C \varepsilon L}{\dfrac{RT}{F}}$ |

**Note**:
$*$ *corresponds to $\gamma$ in Mc Donald's paper* [10]
$**$ *corresponds to $\Gamma$ in Mc Donald's paper* [10] *and to $\delta$ in Emmanuel's paper* [14]
$***$ *for the present case we have assumed that $\chi = G$*

**Introduction:** Passivity refers to the natural ability possessed by metals to form a protective oxide coat on their surfaces by reacting with oxygen present in the air, at normal temperature and pressure. This protective coating, known as the passive layer or the barrier oxide layer, masks the metal from any further exposure to the environment and saves the metallic structure from corrosion. The question is how and why the barrier oxide layer which is initially compact and robust develops instability and breaks down. In the case of porous over-layers on metals much is known about the mechanisms of corrosion when such surfaces are exposed to corrosive media [1] or to the atmosphere [2]. On the other hand the breakdown of compact oxide layers on metals calls for completely new approaches.

Passivity was first observed and reported by Michael Faraday. Several theoretical efforts proposed to model the passive layer and its stability were extensively reviewed [3-5], and the point defect model (PDM), initially proposed in 1981 by McDonald [6,7] happens to be widely accepted till date to model the stability issues related to the passive film on several metals and alloys.

The PDM employs concepts from the physics of point defects in solids and electrochemistry to address the passivation phenomenon by proposing 7 defect reactions [8]. The defect reactions that take place at the metal |metal oxide interface and at the metal oxide | solution interface are indicated in Fig. 1a. While the conjugate reactions 3 and 6, also known as the oxide thickening reactions, addresses the oxide formation on the metal surface, the reaction 7 refers to the dissolution of the passive film at the f/s interface which makes local thinning of the passive film. The rates of thickening and thinning are equal in the steady state. An important deficiency in the PDM that needs correction is that the PDM reaction 3 is not balanced and thereby does not clearly explain the origin of oxygen vacancies. Furthermore, PDM suggests [9] that if the metal vacancies [$V_M^{x-}$] generated by reaction (4) is not fully consumed by its conjugate reaction (1) then the unconsumed vacancies [$V_M^{x-}$] will condense to form voids at the m/f interface. This statement, however, does not obey the well-known coulomb law that like charges will repel each other.

To remedy this situation, Emmanuel introduced the Variant Point Defect Model (VPDM) [4, 5]. As far as the interfacial defect reactions are concerned, the VPDM differs from the PDM only in reaction 3. In the context of the VPDM, throughout this paper, the reaction (3) will be replaced by reaction 3'.

**Fig. 1a:** PDM reactions as enunciated by Mc Donald

| Metal - Metal oxide Interface (Reactions at X = 0) | Metal oxide - Electrolye Interface (Reactions at X = L) |
|---|---|
| **Defect reaction 1** $$m + V_M^{x} \rightarrow M_M + V_m + \chi e^-$$ The metal ion vacancy is consumed) & a vacancy is created in the metal | **Defect reaction 4** $$M_M \rightarrow M^{\delta+}_{(aq)} + V_M^{x} + (\delta-\chi) e^-$$ Metal ion vacancy is created in the metal oxide |
| **Defect reaction 3** $$m \rightarrow M_m + (\chi/2) V_o^{**} + \chi e^-$$ Creation of oxygen vacancy is not clear | **Defect reaction 6** $$V_o^{**} + H_2O \rightarrow O_o + 2H^+$$ Oxygen vacancy is consumed & $H^+$ is generated |

**Defect reaction 7**
$$MO_{\chi/2} + \chi H^+ \rightarrow M^{\delta+}_{(aq)} + (\chi/2) H_2O + (\delta-\chi) e^-$$   $H^+$ interacts with the metal oxide and leads to its dissolution

**Fig.1b:** PDM reactions as modified by Emmanuel under his VPDM

| Metal - Metal oxide Interface (Reactions at X = 0) | Metal oxide - Electrolye Interface (Reactions at X = L) |
|---|---|
| **Defect reaction 1** $$m + V_M^{x} \rightarrow M_M + V_m + \chi e^-$$ The metal ion vacancy is consumed) & a vacancy is created in the metal | **Defect reaction 4** $$M_M \rightarrow M^{\delta+}_{(aq)} + V_M^{x} + (\delta-\chi) e^-$$ Metal ion vacancy is created in the metal oxide |
| **Defect reaction 3'** $$m + q V_m + (\chi/2) O_o \rightarrow MO_{\chi/2} + (\chi/2) V_o^{**} + \chi e^-$$ The vacancy in the metal is consumed & an Oxygen vacancy is created | **Defect reaction 6** $$V_o^{**} + H_2O \rightarrow O_o + 2H^+$$ Oxygen vacancy is consumed & $H^+$ is generated |

**Defect reaction 7**
$$MO_{\chi/2} + \chi H^+ \rightarrow M^{\delta+}_{(aq)} + (\chi/2) H_2O + (\delta-\chi) e^-$$   $H^+$ interacts with the metal oxide and leads to its dissolution

*Kroger-Vink notations used in the defect reactions dealt with above are as follows: m = metal; $V_M^{x-}$ = metal vacancy in the metal oxide; $M_M$ = metal ion in the metal oxide sub-lattice; $V_m$ = vacancy in the metal; $O_o$ = oxygen in the oxygen sub-lattice. $V_o$ = oxygen ion vacancy. (The interstitial reactions represented in the PDM vide equations (2) and (5) are not considered in the present work and hence are left out here. The complete set of PDM reactions can be found in the literature by McDonald [8] and in the articles by Emmanuel [4, 5].*

According to the PDM as well as per the VPDM, the defect reaction 1 that occurs at the M/MO interface, involves the consumption of a metal ion vacancy ($V_M^x$) which was generated at the MO/electrolyte interface vide the defect reaction 4. However, the PDM does not clearly explain the defect reaction 3. On the other hand, the VPDM reaction 3' clearly explains the formation of oxygen vacancies. The vacancy in the metal ($V_m$) is consumed and an oxygen vacancy is created. This reaction is oxide thickening reaction since metal oxide is formed as the product. Furthermore, equation 3' gives a theoretical footing to the well-known empirical relation in corrosion science: the so-called Pilling-Bedworth (PB) ratio. The PB ratio evolves naturally upon rearranging the equation 3' after imposing volume conservation [4, 5]. The other equations (6, 7) are the same for both the PDM and the VPDM, i.e., the oxygen vacancy produced by the reaction 3' interacts with water to produce $H^+$ (defect reaction 6). The $H^+$ reacts with the metal oxide leading to its breakdown (defect reaction 7) and this is the oxide thinning reaction. For more detailed discussion on the PDM and VPDM may be found in [4].

The novelty of the present paper is based upon the idea that condensation is in between oppositely charged vacancies. This constraint satisfies the Coulomb's law. The cation vacancy ($V_M^{x-}$) and the anion vacancy ($V_o^{2+}$), produced respectively by the PDM reaction 4 at the m/f interface and by the VPDM reaction 3' at the f/s interface diffuse into the bulk of the oxide and condense to form voids.

**The Mathematical Model**: The geometry of the system modelled is in Fig. 3 There are two interfaces: (a) the region x=0 where the metal oxide is in contact with the metal is known as the metal | metal oxide interface, and (b) the region x=L where the metal oxide is in contact with the electrolyte is known as the metal oxide | electrolyte interface. 'L' is the thickness of the metal oxide film which is of the order of a few nanometers. Initially we assume that the barrier oxide is compact i.e., there are no pores. Defect reactions occurring at the metal | barrier oxide interface and at the barrier oxide | electrolyte interface lead to the formation of cation vacancies ($V_M^{x-}$) and anion vacancies ($V_o^{2+}$) which diffuse and migrate in opposite directions across the oxide film.

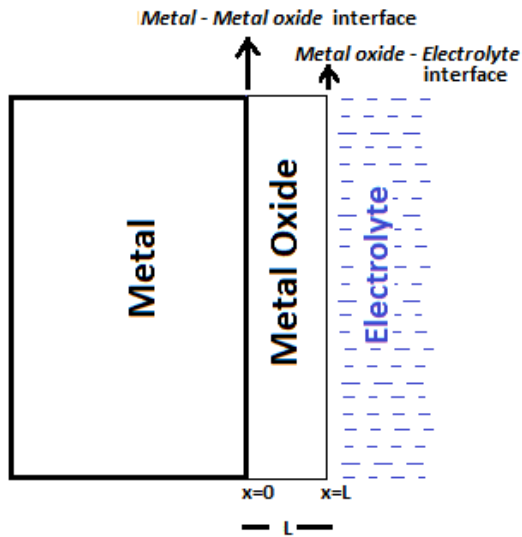

**Fig. 3**: The geometry of the system studied

Therefore we need two partial differential equations (PDEs), one to address the transport of cation vacancies and the other for anion vacancies. The oppositely charged vacancies produced at the m/f and f/s interfaces are driven into the bulk of the oxide by concentration and electric field gradients where they interact with each other to form a defect pair (DP). For instance in the case of FeO the defect pair could be represented as [$V_{Fe}^{2-}V_o^{2+}$]. Though Fe is the cation, the corresponding defect will be negatively charged (for reasons of charge neutrality) and vice versa. Several such defect pairs cluster together to form a void.

Besides the diffusion and migration terms in the PDEs we need a source/sink term to account for the change of the concentrations of the anion and cation defects in the oxide due to the formation of defect pairs which further cluster to form voids. The physical picture behind the model is that the defect pairs that are generated in the oxide film join together and grow into voids by a nucleation and growth mechanism. These voids grow from pre-existing sites – the so-called active sites in nucleation and growth theories – which may energetically favor the accumulation of defect pairs.

**The PDEs are:**

$$\frac{\partial C_A}{\partial t} = D_{A_{eff}} \frac{\partial^2 C_A}{\partial x^2} + \frac{Z_A * D_{A_{eff}} * E * L}{\left(\frac{RT}{F}\right)} \left(\frac{\partial C_A}{\partial x}\right) - \left\{\exp\left[-\frac{4}{3}\pi N_o R^3(x,t)\right] * 4\pi N_o R^2(x,t)\right\} * \{(k_{vf} C_A C_C - k_{vb})\} \quad \rightarrow \quad (1)$$

$$\frac{\partial C_C}{\partial t} = D_{C_{eff}} \frac{\partial^2 C_C}{\partial x^2} + \frac{Z_C * D_{C_{eff}} * E * L}{\left(\frac{RT}{F}\right)} \left(\frac{\partial C_C}{\partial x}\right) - \left\{\exp\left[-\frac{4}{3}\pi N_o R^3(x,t)\right] * 4\pi N_o R^2(x,t)\right\} * \{(k_{vf} C_A C_C - k_{vb})\} \quad \rightarrow \quad (2)$$

$D_{A_{eff}}$ and $D_{C_{eff}}$ are effective diffusion coefficients which recognize the presence of voids in the oxide film and these in fact depend on the x- and t- dependent void radius $R(x,t)$ [See the Note at the end of the paper]. The first and second terms on the RHS are respectively the diffusion and migration terms while the third term is the source/sink term accounting for the consumption or the production of anion and cation defects at the space-time point (x,t). The third term is a product of two factors in curly brackets. The second factor is the local flux of the elementary defect pairs which go into or come out of a void of radius $R(x,t)$. The first factor is the area per unit volume provided by the assembly of voids present in a thin slice of the oxide film at (x, t). The actual form of this factor follows from the Avrami's theorem [ 10 ] originally developed for quantifying the extent of phase transformations in solid state reactions and it further incorporates the more recent connection established by Emmanuel between the volume of a random agglomerate and its bounding area [ 11 ].

**Boundary conditions (BCs):** Each PDE has two boundary conditions, one at x=0 and the other at x=L. These are given in the Table below.

| Dimensional form of the BCs |
|---|
| $-D_{A\,eff}\frac{\partial C_A}{\partial x} + k_A C_{A(0,t)}\,\|_{x=0} = k_3$ |
| $-D_{A\,eff}\frac{\partial C_A}{\partial x} + k_A C_{A(1,t)}\,\|_{x=L} = k_6 C_{A(1,t)}$ |
| $-D_{C\,eff}\frac{\partial C_C}{\partial x} + k_c C_{C(0,t)}\|_{x=0} = -k_1 C_{C(0,t)}$ |
| $-D_{C\,eff}\frac{\partial C_C}{\partial x} + k_c C_{C(1,t)}\|_{x=L} = -k_4$ |

These boundary conditions capture the defect reactions at the two interfaces. The RHS of these BCs have the diffusion and migration fluxes while their LHS contain the rate constants $k_1, k_3, k_4$ and $k_6$ of the four defect reactions considered in the present model. These rate constants depend on the applied voltage (V), oxide film thickness (L) and the pH. Macdonald parameterized these as [12]:

$k_i = k_i^o\, e^{a_i V + b_i L + c_i pH}$

Where $k_i^o$ the standard is rate constant of *the i-th* reaction. $a_i, b_i\ and\ c_i$ are parameters defined in Table below.

| Defect reaction index | $a_i$ (V$^{-1}$) | $b_i$ (cm$^{-1}$) | $c_i$ |
|---|---|---|---|
| 3' | α₃ (1-α) χ g | -α₃ χ ε g | -α₃ β χ g |
| 6 | 2 α₆ α g | | 2 α₆ β g |
| 1 | α₁ (1-α) χ g | -α₁ χ ε g | -α₁ β χ g |
| 4 | α₄ α G g | | α₄ β G g |

Macdonald et al had evaluated the values of these standard rate constants from some base rate constants. However the values of the base rate constants as published by Macdonald group shows large deviations from one paper [12] to the other [13]. These deviation is as large as 30 orders of magnitude. Even if the systems under study are completely different, such huge deviations are unacceptable. Hence we sought to evaluate these standard rate constants from a heuristic analysis which is outlined in the Appendix I.

**The ODE:**

The model will be incomplete without a law for the distribution and growth of the voids that nucleate inside the oxide film when the anion and cation defects collide to form a neutral defect pair (DP). We already discussed that defects generated at the M | BL and at the BL | El interfaces move into the bulk of the oxide through diffusion and migration effects. Oppositely charged vacancies combine to form defect pairs. The DPs are captured by the nucleating sites available in the metal oxides, to form voids and the radius of the void grows. In conformity with standard nucleation theories [14] we assume the presence of active sites in the oxide film at which the voids nucleate and grow by further addition of DPs as indicated below:

$$n\,DP\ + DP\ \rightarrow (n+1)DP$$

The flux corresponding to the addition of these DPs to the void is given by the equation:

$$Flux = k_f C_A C_C - k_b \quad \rightarrow (3)$$

This is the radial flux in or out of the void and depends on x and t through the x and t dependence of the anion and cation concentrations.

By an elementary analysis [See Appendix II] one can show that the radius $R(x,t)$ of a void at the location $x$ follow the ordinary differential equation (ODE):

$$\frac{\partial R}{\partial t} = V_m (k_f C_A C_C - k_b) \quad \rightarrow (4)$$

**Initial conditions for the PDEs and the ODE:** we assume that the barrier oxide has initially some defects. These defects are either naturally generated during the oxide growth or introduced later from the environment (e.g. neutron flux in a nuclear reactor). Thus the initial conditions for the two PDEs are: $C_A(x,0) = C_A^o$ and $C_C(x,0) = C_C^o$. The ODE needs one initial condition $R(x,0) = R_o$.

**Non-dimensional forms of the governing equations and the boundary conditions:**

**It is convenient to work with non-dimensional forms. Table below** lists the dimensional variables, the scaling factor and the corresponding non-dimensional forms.

| Dimensional variables in the VPDM | Scaling factor | Non-dimensional form |
|---|---|---|
| **x** (space) | Division by $L$ (thickness of the barrier oxide) | $x' = \frac{x}{L}$ |
| **R** (Radius of the void) | Division by $L$ | $R'(x,t) = \frac{R(x,t)}{L}$ |
| **t** (time) | Division by $\frac{L^2}{D}$ (D is the average diffusion coefficient defined as $D = (\frac{D_C + D_A}{2})$) | $t' = \frac{t}{\frac{L^2}{D}}$ |
| **C$_c$** (Concentration of the cation vacancy) | Division by the initial cation defect concentration $C_C^o$ | $C_C' = \frac{C_C}{C_C^o}$ |
| **C$_A$** (Concentration of the anion vacancy) | Division by the initial anion defect concentration $C_A^o$ | $C_A' = \frac{C_A}{C_A^o}$ |
| **$N_o$** (Number of nucleating sites. $N_o$ may vary between milli- to micro-molar concentration) | Multiplication by $L^3$ | $N_o' = N_o * L^3$ |
| **CP** (defined as the product of C$_A$ and C$_C$) | Division by $C_A^o C_C^o$ | $CP' = \frac{C_A}{C_A^o} \frac{C_C}{C_C^o}$ |
| $D_{C_{eff}}$ (corrected value of the diffusion coefficient for the cation vacancy) $D_{C\,eff} = D_C * \exp(-\frac{4}{3}\pi \gamma N_o' R^3(x,t))$ | Division by the average diffusion coefficient D | $D_C' = \frac{D_{C_{eff}}}{D}$ |
| $D_{A_{eff}}$ (corrected value of the diffusion coefficient for the anion vacancy) $D_{A\,eff} = D_A * \exp(-\frac{4}{3}\pi \gamma N_o' R^3(x,t))$ | Division by the average diffusion coefficient D | $D_A' = \frac{D_{A_{eff}}}{D}$ |
| $k_3$ (Rate constant for the defect reaction 3' of the VPDM) | Multiplication by $\frac{L}{D_A * C_A^o}$ | $K_3' = k_3 * \frac{L}{D_A * C_A^o}$ |
| $k_6$ (- do - for defect reaction 6) | Multiplication by $\frac{L}{D_A}$ | $K_6' = k_6 * \frac{L}{D_A}$ |
| $k_1$ (- do – for defect reaction 1) | Multiplication by $\frac{L}{D_C}$ | $K_1' = k_1 * \frac{L}{D_C}$ |
| $k_4$ (- do – for defect reaction 4) | Multiplication by $\frac{L}{D_C * C_C^o}$ | $K_4' = k_4 * \frac{L}{D_C * C_C^o}$ |

Using these non-dimensional variables the non-dimensional governing equations become:

**PDEs:**

$$\frac{\partial C_A'}{\partial t} = D_A' \frac{\partial^2 C_A'}{\partial x'^2} + \frac{Z_A * E * L}{\left(\frac{RT}{F}\right)} D_A' \left(\frac{\partial C_A'}{\partial x'}\right) - \exp\left[-\frac{4}{3}\pi N_o' R'^3\right] * R'^2 * t_{RA} * [C_A' C_C' - CP'] \quad \rightarrow (5)$$

$$\frac{\partial C_C'}{\partial t'} = D_C' \frac{\partial^2 C_C'}{\partial x'^2} + \frac{Z_C * E * L}{\left(\frac{RT}{F}\right)} D_C' \left(\frac{\partial C_C'}{\partial x'}\right) - \exp\left[-\frac{4}{3}\pi N_o' R'^3\right] * R'^2 * t_{RC} * [C_A' C_C' - CP'] \quad \rightarrow (6)$$

**ODE:**

$$\frac{dR'}{dt'} = K_G (C_A' C_C' - CP')$$

The non-dimensional forms of initial and boundary conditions are given in the Table below.

| PDE | Initial condition | Interface | Defect reaction | Boundary conditions (Non-Dimensional form) |
|---|---|---|---|---|
| PDE equation (5) | $C_A = C_A^o$ | x=0 | 3' | $-\frac{\partial C_A'}{\partial x'} + K_A C_A \|_{x'=0} = K_3' \exp\left[\frac{4}{3}\pi \gamma N_o' R^3(0,t)\right]$ |
| | | x=L | 6 | $-\frac{\partial C_A'}{\partial x'} + K_A C_A \|_{x'=1} = K_6' \exp\left[\frac{4}{3}\pi \gamma N_o' R^3(1,t)\right] C_A (1,t)$ |
| PDE equation (6) | $C_C = C_C^o$ | x=0 | 1 | $-\frac{\partial C_C'}{\partial x'} + K_C C_C \|_{x'=0} = -K_1' \exp\left[\frac{4}{3}\pi \gamma N_o' R^3(0,t)\right] C_C (0,t)$ |
| | | x=L | 4 | $-\frac{\partial C_C'}{\partial x'} + K_C C_C \|_{x'=1} = -K_4' \exp\left[\frac{4}{3}\pi \gamma N_o' R^3(1,t)\right]$ |

Results and Discussion:

Some initial results are presented and discussed here. More results will be presented elsewhere. The governing equations were solved using MAPLE 2018. The cation and anion defects are generated at the opposite ends of the oxide layer, move across the film and coalesce to produce the void. Hence the void radius should be minimum at the interfaces (x=0 and x=1) and maximum at x=0.5. The anion defect concentration decreases as it should from the metal-oxide junction and the cation defect concentration decreases from the oxide-electrolyte junction as expected.

In the implementation of the present model, two physical constraints need to be kept in mind: 1] The void radius should not be allowed to decrease below its initial radius to avoid unphysical negative radii and 2] The voids should not be allowed to grow beyond the spatial boundaries of the oxide film. These two constraints are incorporated in equation (4).

The results presented below show only typical trends and we have not yet tuned the model parameters to correspond to real material systems which is to be taken up. These figures are self-explanatory and captions are not provided. The quantities on the figure axes are non-dimensional except for the current density which is in $A\ cm^{-2}$.

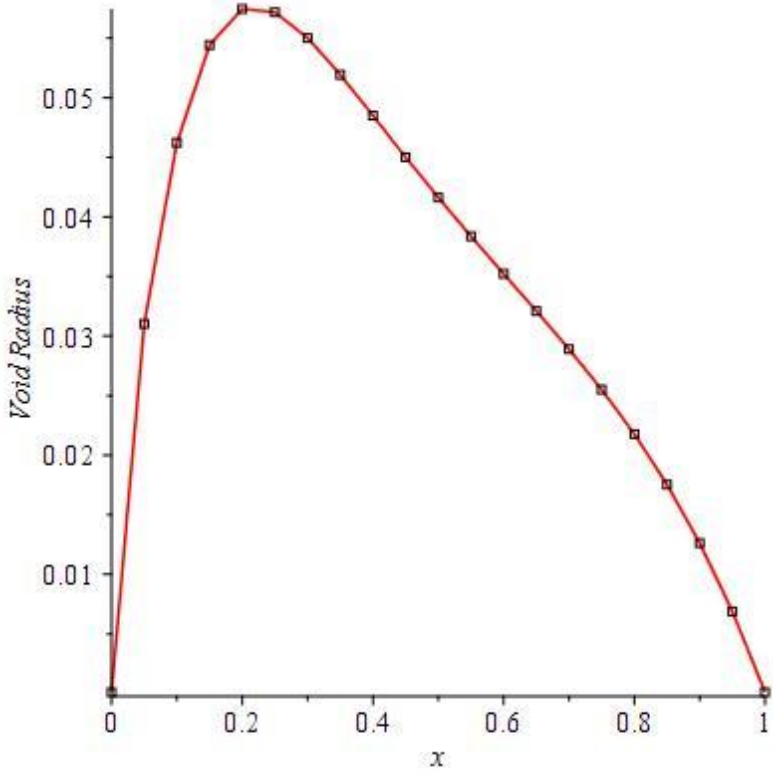

Figure-1 (for t=600)

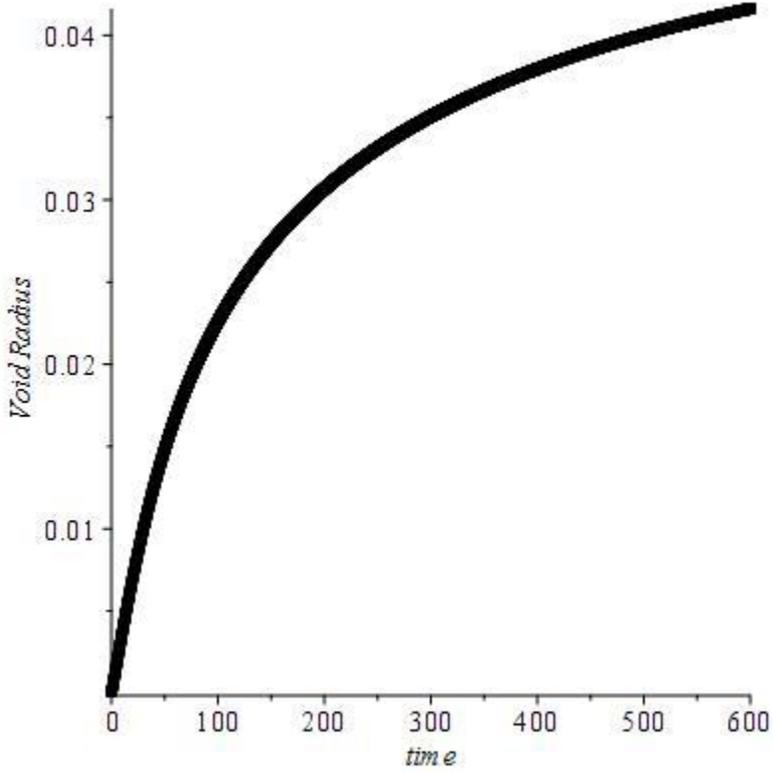

Figure-2 (for x=0.5)

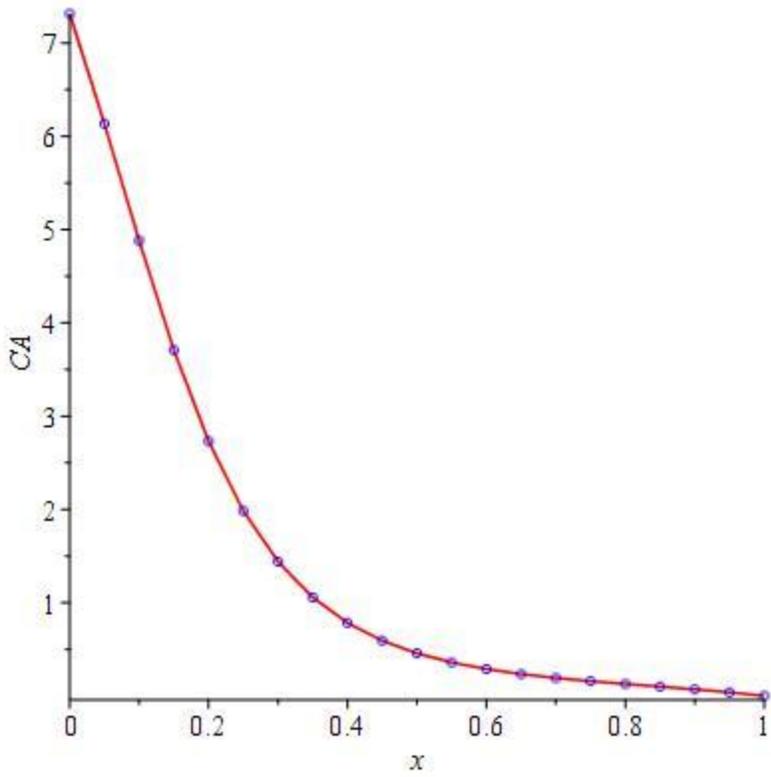

Figure-3 (for t=600)

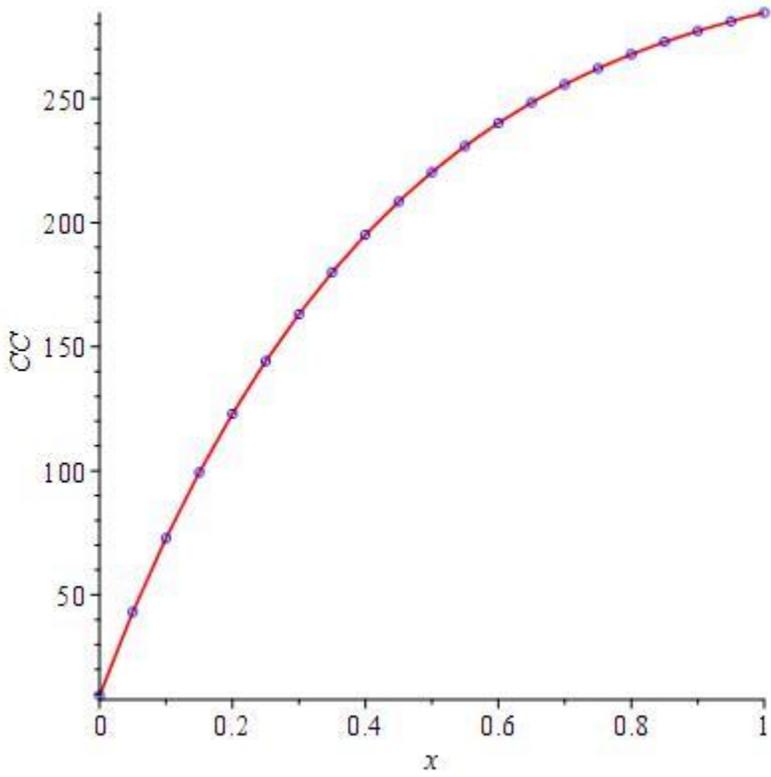

Figure-4 (for t=600)

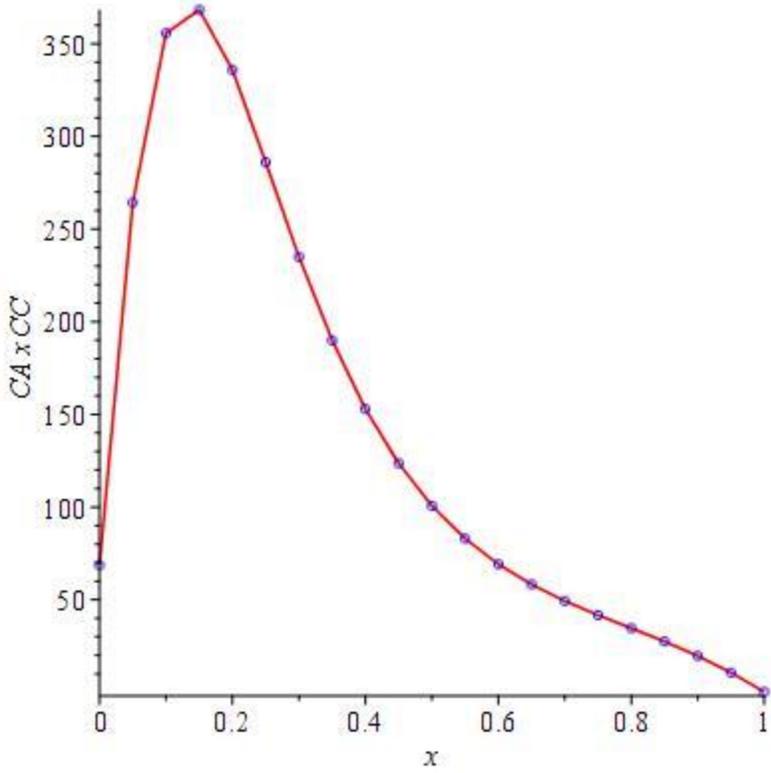

Figure-5 (for t=600)

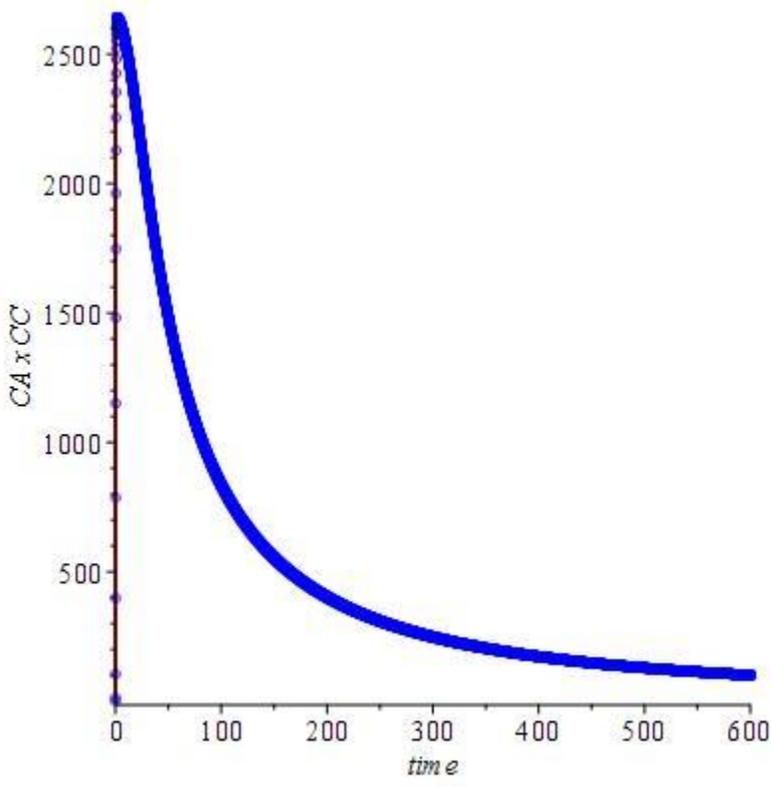

Figure-6

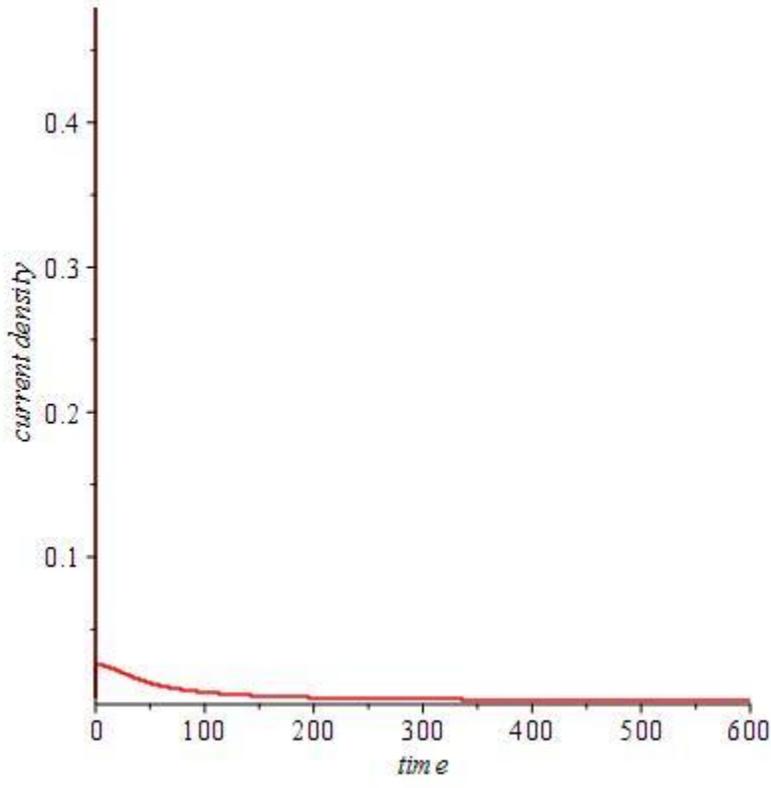
Figure-7

Appendix I

Consider first the standard rate constant $k_1^o$ for the defect reaction 1. Its dimension is $\frac{m}{s}$. From a careful study of reaction 1, one can see that this reaction corresponds to a metal atom in the metal at the metal|metal oxide interface hopping into the metal ion vacancy (created through the defect reaction 4) in the adjoining metal oxide. The hopping distance can be of the order of a few Å (say ~ 10 Å). As hopping is similar to diffusion we may estimate $k_1^o$ by dividing a typical value of the solid-state diffusion coefficient like $10^{-13}\frac{m^2}{s}$ by a typical hopping distance like $10^{-9}$ m to obtain $k_1^o = 10^{-4}\frac{m}{s}$.

Consider next $k_4^o$ : This has the dimension of flux. The steady state passive current density is of the order of $\frac{1\mu A}{cm^2}$ i.e., $\frac{10^{-2}A}{m^2}$ in SI units.

Current density = n x Faraday x Flux

Assuming n=1

Flux = $\frac{current\ density}{Faraday}$ = $10^{-7}\frac{mole}{m^2 sec}$

A reasonable estimate of $K_4°$ can thus be $10^{-7}\frac{mole}{m^2 sec}$

Concerning the order of magnitudes for $k_3^o$ and $k_6^o$ , these standard rate constants have the dimensions of $\frac{mole}{m^2 sec}$ and $\frac{m}{s}$ respectively. Hence we have assumed that they possess values of the same order of magnitude as that for $k_4^o$ and $k_1^o$ respectively.

---

**Appendix II**

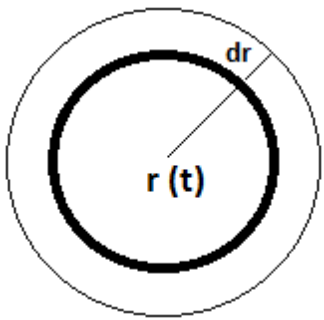

Consider the above single void of radius r(t) at time t and its differential increment dr by the addition DPs in time dt. Clearly $\frac{4\pi r^2 dr}{V_m}$ is the number of moles of DPs added to the void that should equal that given by $Flux * 4\pi r^2 * dt$.

Equate the two to obtain

$$\frac{\partial r}{\partial t} = V_m \left( k_f C_A C_C - k_b \right)$$

---

**Note on the effective diffusion coefficient $D_{eff}$**

As the voids grow and porosity develops in the oxide film, the value of the diffusion coefficient (D) of any species entering as an input into the model has to be corrected. The corrected value, known as the effective diffusion coefficient $D_{eff}$ is related to uncorrected D by the empirical relation:

$$D_{eff} = D(1 - \varepsilon_p)^\gamma$$

$\varepsilon_p$ represents the porosity, which is the ratio of the void volume to the total volume. $\gamma$ is the tortuosity.

From Avrami's relation we know: $\varepsilon_p = 1 - \exp\left[-\frac{4}{3}\pi N_o R^3(x,t)\right]$

Therefore $\qquad D_{eff} = D * \exp\left[-\frac{4}{3}\pi N_o \gamma R^3(x,t)\right]$

---

**Acknowledgement:** K. Ragavendran acknowledges support from IMSc through an associateship program and leave of absence from Kalasalingam Academy of Research and Education. Bosco Emmanuel thanks G. Bakaran (IMSc) for his support of this work by SERB grant no. SB/DF/005/2014 of Department of Science and Technology (New Delhi). The authors acknowledge interesting discussions with G. Baskaran and IMSc for hospitality.

**References**:

[1] M.S. Venkatraman, I.S. Cole, Bosco Emmanuel Electrochim. Acta 56 (2011) 7171;

M.S. Venkatraman, I.S. Cole, Bosco Emmanuel Electrochim. Acta 56 (2011) 8192;

I.G. Bosco, M.S. Venkatraman, I.S. Cole, Bosco Emmanuel ECS Trans. 35 (2011)1;

I.G. Bosco, I.S. Cole, Bosco Emmanuel J.Electroanal.Chem. 727 (2014)68;

[2] D. Sherwood, M.V. Reddy, I.S. Cole, Bosco Emmanuel J.Electroanal.Chem. 725 (2014)1;

D. Sherwood, Bosco Emmanuel, I.S. Cole J.Electrochem.Soc. 163 (2016) C675;

   D. Sherwood, Thesis, March 2017

[3] D.D. Macdonald, Russ. J. Electrochem. 48 (2012) 235.

[4] A. Veluchamy, D. Sherwood, B. Emmanuel, Ivan S Cole, J. Electroanalytical chemistry 785 (2017) 196-215.

[5] B. Emmanuel, https://arxiv.org/ftp/arxiv/papers/1208.1208.1096.pdf

   B. Emmanuel, https://arxiv.org/ftp/arxiv/papers/1304/1304.3227.pdf

[6] C.Y. Chao, L.F. Lin, D.D. Macdonald, J. Electrochem. Soc. 128 (1981) 1187.

[7] D.D. Macdonald, J. Electrochem. Soc. 139 (1992) 3434.

[8] D.D. Macdonald, S.R. Biaggio, H. Song, J. Electrochem. Soc. 139 (1992) 170.

[9] S.J. Ahn, H.S. Kwon, D.D. Macdonald, J. Electrochem. Soc. 152 (2005) B482.

[10] B. Emmanuel, J. Chem. Soc. Faraday Trans.I, 1981, 77, 483-495.

[11] B. Emmanuel, Phys.Rev.E, 1995, 52, 4681-4684.

[12] D.D. Macdonald, J. Electrochem. Soc. 153 (2006) B 213.

[13] S.J. Ahn, H.S. Kwon, D.D. Macdonald, J. Electrochem. Soc. 152, (2005), B482.

[14] Bosco Emmanuel, Thesis, IISc, 1982